\documentclass[prb,twocolumn,showpacs,superscriptaddress]{revtex4}

\usepackage{graphicx}
\usepackage{bm}
\usepackage{amsmath,amssymb}
\usepackage{epsf}
\newcommand{\beq }{\begin{eqnarray}}
\newcommand{\eeq }{\end{eqnarray}}
 
\begin{document}
\title
{Chaotic Transport in the Symmetry Crossover Regime 
with a Spin-orbit Interaction}


\author{Keiji Saito}
\affiliation{Department of Physics, Graduate School of Science,
University of Tokyo, Tokyo 113-0033, Japan}
\affiliation{CREST, JST, 4-1-8 Honcho Kawaguchi, 
Saitama, 332-0012, Japan}

\author{Taro Nagao}
\affiliation{Graduate School of Mathematics,
Nagoya University, Chikusa-ku, Nagoya 464-8602, Japan} 

\date{\today}

\begin{abstract}
We study a chaotic quantum transport in the presence of 
a weak spin-orbit interaction.
Our theory covers the whole symmetry crossover regime between time-reversal 
invariant systems with and without a spin-orbit interaction. This 
situation is experimentally realizable when the spin-orbit interaction 
is controlled in a conductor by applying an electric field. 
We utilize a semiclassical approach 
which has recently been developed. In this approach, the non-Abelian nature of 
the spin diffusion along a classical trajectory plays a crucial role. 
New analytical expressions with one crossover parameter are 
semiclassically derived for the average conductance, conductance 
variance and shot noise. Moreover numerical results on a random matrix 
model describing the crossover from the GOE (Gaussian Orthogonal 
Ensemble) to the GSE (Gaussian Symplectic Ensemble) are compared 
with the semiclassical expressions.

\end{abstract}

\maketitle
\section{Introduction}

A chaotic quantum transport of an electron in a cavity is caused 
by either implanted impurities or bumpy boundaries, and provides 
directly measurable {\em quantum signatures of chaos}\cite{haakebook}, 
such as the conductance variance. Universal aspects of a chaotic 
transport have been investigated by means of the random matrix theory 
(RMT)\cite{beenakker97}. In the RMT, quantum systems are classified 
into symmetry classes. A chaotic system with 
time-reversal symmetry is described by the Gaussian orthogonal ensemble 
(GOE). When the time-reversal symmetry is broken by applying a magnetic 
field, the Gaussian unitary ensemble (GUE) becomes a suitable model. 
If a system with time-reversal symmetry has a spin-orbit interaction, one 
needs to employ the Gaussian symplectic ensemble (GSE).
\par 
We consider the case that two leads are attached to a cavity 
and the number of the 
lead channels are $N_1$ and $N_2$. An electron transport in the cavity is 
described by the scattering matrix\cite{landauer,buttiker}. Replacing the 
scattering matrix by a random matrix, the RMT phenomenologically predicts 
the average conductance $G$, conductance variance ${\rm Var} G$, 
and shot noise $P$ at zero temperature as\cite{rmt,savin}
\beq
\frac{G}{G_0} 
\!\!&=&\! \frac{2 N_1 N_2}{N -1 + \frac{2}{\beta}}  \nonumber \\
\!\!&=&\!\! \frac{2 N_1 N_2}{N} \left\{ 1 + \frac{1-\frac{2}{\beta}}{N}
+ \left( \frac{1-\frac{2}{\beta}}{N} \right)^2 \right\} 
\nonumber \\ & & + O\left( \frac{1}{N^2} 
\right),~~ \label{rdm_conductance}  
\eeq
\beq
\frac{{\rm Var} G}{G_0^2} \!\! &=&\!\! 
\frac{8 N_1 (N_1 - 1 + \frac{2}{\beta}) N_2 (N_2 - 1 + 
\frac{2}{\beta})}{ 
\beta (N-2+ \frac{2}{\beta}) (N-1 +\frac{2}{\beta})^2 (N-1 + \frac{4}{\beta}) 
} \nonumber \\
\!\!&= &\!\! \frac{8 N_1^2 N_2^2}{\beta N^4} + O\left(\frac{1}{N}\right),
\label{rdm_conductance_fluc} 
\eeq
\beq
\frac{P}{P_0} \!\!&=&\!\! 
\frac{2N_1 (N_1 - 1 + \frac{2}{\beta} ) N_2 (N_2 - 1 + \frac{2}{\beta})}{
(N -2 + \frac{2}{\beta}) (N -1 +\frac{2}{\beta} )(N-1+\frac{4}{\beta} )
} \nonumber \\
\!\! &=&\!\!
\frac{2 N_1^2 N_2^2}{N^3} + \left( \frac{4}{\beta} 
-2 \right) \frac{N_1 N_2}{N^4}
(N_1 - N_2 )^2 \nonumber \\ & & + O\left(\frac{1}{N}\right)  
\label{rdm_shot_noise}
\eeq
with $N = N_1 + N_2$. Here $\beta=1,2$, and $4$ correspond to the GOE, GUE, 
and GSE symmetry classes, respectively. These expressions include the 
contributions from the spin degrees of freedom\cite{comment}. The constants 
$G_0$ and $P_0$ are $G_0 = {e^2 /(\pi \hbar)}$ and $P_0 ={2 e^3 |V| /(\pi\hbar)}$, respectively, where $e$ is the unit electric charge and $V$ is 
the bias voltage. If $N_1$ is equal to $N_2$ and very large, the leading 
term of the shot noise is insensitive to a change of the symmetry. 
\par
When a very weak magnetic field is applied to the cavity, 
the time-reversal symmetry is only partially broken. In this case, a crossover 
from the GOE to GUE is observed. This GOE-GUE crossover regime can also be 
analyzed by a parametric RMT model, and analytic predictions describing a 
chaotic quantum transport are known\cite{weidenmuller}. Recently, a chaotic 
transport in the GSE-GUE crossover regime was also studied within the RMT 
framework\cite{kumar_pandey}. In this regime, a very weak magnetic field 
breaks the time-reversal symmetry of a system with a spin-orbit interaction.
\par
The aim of this paper is to study another case, the crossover from the GOE 
to GSE, in which the system has a very weak spin-orbit interaction 
preserving the time reversal symmetry. In the experimental point of view, 
the GOE-GSE crossover can be realized, if the spin-orbit interaction 
(or Aharonov-Casher effect) is controlled by applying an electric field in a 
chaotic conductor\cite{rashba,nitta}. In this case, a parametric RMT 
model is also known\cite{brouwer1,brouwer2}, and the diagrammatic 
perturbation theory has been used to evaluate some transport 
properties\cite{brouwer3,cremers,beri1,beri2}. Here we 
employ a semiclassical approach which has recently been 
developed\cite{bolte2, richter1,richter2,braun,haake,bolte3}. 
In a semiclassical evaluation, the transmission 
amplitude is treated by the path-integral method, where all the classical 
paths must in principle be taken into account. However recent studies 
clarified that almost the same but partially time reversed pairs of 
classical trajectories contributed to the conductance\cite{aleiner,
takane,rs02,sr_pair}, so that the calculation was greatly simplified. 
Then it was shown that the semiclassical approach could precisely reproduce 
the RMT predictions (\ref{rdm_conductance})-(\ref{rdm_shot_noise})\cite{braun,
haake,bolte3}. Moreover, when a similar approach is applied to the parametric 
spectral correlations in the GOE-GUE, GUE-GUE, GOE-GOE and GSE-GSE regimes, 
it can also reproduce the RMT predictions\cite{saito_nagao,nagao07,kuipers,nagao_saito}.
\par
Thus the semiclassical approach has become a practical tool to find 
a new prediction, even if the RMT analysis is difficult. As this approach was 
already applied to the parametric spectral correlations in the GOE-GSE 
crossover regime\cite{nagao_saito}, we naturally expect that it can 
be used in the analysis of a transport. 
\par
Considering the non-Abelian nature of the spin diffusion along the 
classical trajectories, we extend the semiclassical technique to 
derive analytic expressions for the transport properties. Our results 
on the average conductance, conductance variance and shot noise are 
given in eqs. (\ref{result_g}), (\ref{result_varg}), and (\ref{result_P}). 
The crossover from the GOE to GSE is controlled by one parameter 
depending on the diffusion constant of the spin. The GOE and GSE 
results are reproduced in the limiting cases of the parameter. 
\par
This paper is organized in the following way.  In Sec. II, 
a semiclassical expression of the transmission amplitude is presented. 
We put a stress on the statistical aspects of the expression. 
In Sec. III, using the semiclassical expression, we calculate 
the average conductance, conductance variance and shot noise.
In Sec. IV, these results are compared with numerical 
calculations on a random matrix model. We finally summarize 
the paper in Sec. V.

\section{Semiclassical expression of the transmission amplitude}

The semiclassical theory employs the transmission amplitude
$t_{a_1, a_2}$, which represents the propagator of a wave packet 
from the channel $a_1$ in one lead to $a_2$ in another lead.
Bolte and Keppeler derived a semiclassical expression of the 
transmission amplitude with spin variables\cite{bolte1}:
\beq
t_{a_1, a_2} &\sim & \sqrt{\frac{2}{T_H}}
\sum_{\alpha : a_1 \to a_2 } {\cal A}_{\alpha } \Delta_{\alpha} 
e^{i S_{\alpha}/ \hbar}, \label{transmission}
\eeq
where $T_H$ is the Heisenberg time $T_H= \frac{\Omega (E) 
}{(2\pi \hbar )^{f-1}}$. Here $\Omega (E)$ is the 
phase volume density including spin degrees of freedom 
at the energy $E$, and $f$ is the spacial dimension. 
Throughout this paper, we study the two dimensional case 
$f=2$. Two leads are assumed to have $N_1$ and $N_2$ channels, 
i.e. $a_1=1,2,\cdots, N_1$ and $a_2 =1,2,\cdots, N_2$.
The classical action of the orbit $\alpha$ is 
$S_{\alpha}=\int_{\alpha} {\bm p}\cdot d {\bm q}$, where 
${\bm q}$ and ${\bm p}$ are the position and momentum 
variables.

The stability amplitude is decoupled into two factors ${\cal A}_{\alpha}$ 
and $\Delta_{\alpha}$. The first factor ${\cal A}_{\alpha}$ accounts 
for the stability in the position and momentum space, and the second factor 
$\Delta_{\alpha}$ originates from the spin dynamics. Both ${\cal A}_{\alpha}$ 
and $\Delta_{\alpha}$ are uniquely determined, when the classical trajectory 
$\alpha$ governed by the microscopic Hamiltonian is specified. However their 
statistical behavior is independent of the details of the trajectory. 
As discussed in Refs.\onlinecite{braun,haake} and \onlinecite{baranger}, 
the stability amplitude 
${\cal A}_{\alpha}$ is related to the survival probability in the 
chaotic cavity. Postulating an ergodic motion in the position and 
momentum space, we obtain the following sum rule 
\beq
\sum_{\alpha:a_1\to a_2 } |{\cal A}_{\alpha}|^2 &=& 
\int_{0}^{\infty} dT e^{-(2N/T_H ) T}  
, \label{A_decay}
\eeq
where $T_H/2N$ ($N=N_1 + N_2$) can be regarded as the dwell 
time inside the cavity, so that the inverse is the escape rate. 
The escape rate is related to the position and momentum 
variables and is unrelated to the spin variables. Hence the dwell 
time should be $T_H/2N$ rather than $T_H/N$. The spin matrix 
$\Delta_{\alpha}$ is defined as  
\beq
\Delta (t) &=& e^{i \phi(t) \sigma_z /2 }
\, e^{i \theta (t) \sigma_x  / 2 }
\, e^{i \psi(t) \sigma_z / 2 }
\eeq
along a trajectory $\alpha$, where ${\bm \sigma}=(\sigma_x , 
\sigma_y , \sigma_z)$ consists of the Pauli matrices.
The time evolution of the Euler angles $(\psi(t), \theta(t),\phi(t))$ is 
microscopically determined by the Schr\"{o}dinger equation 
\beq
i\hbar \frac{\partial}{\partial t} \Delta (t) =
{\cal H} \Delta (t),  \label{effective}
\eeq
where ${\cal H}$ is the effective Hamiltonian which 
describes the spin-orbit interaction. We assume that 
the spin dynamics is subdominant in the semiclassical 
limit. That is, the dynamics of the position and momentum 
variables are determined by the spacial Hamiltonian without 
spin degrees of freedom, while the spin is influenced by 
the momentum motion via the spin-orbit interaction. 
The effective Hamiltonian in (\ref{effective}) describes 
such a subdominant dynamics of the spin variables.
A similar hierarchical structure has successfully been 
employed to analyze the GOE-GUE crossover regime\cite{haake,
saito_nagao,nagao07,kuipers}: the resulting physical quantities 
are in agreement with the corresponding RMT expressions. 
An RMT prediction for the parametric spectral form factor 
in the GSE symmetry class was also reproduced in a similar 
way\cite{nagao_saito}. 

Since bumpy boundaries of a cavity induce a chaotic 
behavior in the position and momentum variables, 
the momentum effectively plays a role of a 
stochastic magnetic field\cite{nagao_saito}. 
Then the effective Hamiltonian which describes the 
time evolution of the spin can be written as
\beq
{\cal H} = \gamma_{\rm so} \, {\bm h} \cdot \left( 
\frac{\hbar}{2} {\bm \sigma} \right),
\eeq
where $\gamma_{\rm so}$ is the coupling constant 
and ${\bm  h}=(h_x(t),h_y(t),h_z(t))$ is the effective stochastic 
magnetic field. We assume that the classical spin undergoes a Brownian 
motion on the Bloch sphere\cite{HC} due to the stochastic magnetic 
field satisfying
\beq
\langle\!\langle h_{\alpha}(t) h_{\alpha '} (t' ) \rangle\!\rangle &=& 2 
{\cal D} \delta_{\alpha, \alpha '} \delta (t - t' ),~~~~~\alpha , 
\alpha ' =x,y,z, ~~~~~
\eeq
where the brackets $\langle\!\langle ... \rangle\!\rangle$ denote an average 
over the stochastic process of the magnetic field and ${\cal D}$ is the 
diffusion constant. Then the probability 
density function of the Euler angle $P(\psi , \theta , \phi )$ (with 
the measure $\sin\theta d\psi d\theta d\phi$) obeys the Fokker-Planck equation
\beq
\frac{\partial P}{\partial t} = \gamma_{\rm so}^2 {\cal D} {\cal L} P ,
\eeq
where ${\cal L}$ is the Laplace-Beltrami operator
\beq
{\cal L} &=&
\frac{1}{\sin\theta} \frac{\partial}{\partial \theta} \sin \theta
\frac{\partial}{\partial \theta} \nonumber \\
&+& \frac{1}{\sin^2 \theta}\left( 
\frac{\partial^2}{\partial \psi^2}
+ \frac{\partial^2}{\partial \phi^2}
- 2 \cos\theta \frac{\partial^2}{\partial\psi \partial \phi}
\right). 
\eeq
This stochastic dynamics can exactly be analyzed by using 
Wigner's $D$ function \cite{landau_lifshitz}. Let us suppose that 
the initial Euler angles at $t = 0$ are $\omega' =(\psi' ,\theta', 
\phi')$. Then the conditional probability to 
find the angles at $\omega = (\psi , \theta, \phi)$ after 
time $t$ is
\begin{eqnarray}
g ( \omega ; t | \omega ' )
&=& \sum_{j=0}^{\infty}
\sum_{m=-j}^j\sum_{n=-j}^j \frac{2 j + 1}{32 \pi^2 }
\nonumber \\ 
& \times & D_{m,n}^{j} (\psi , \theta , \phi ) 
\{ D_{m,n}^j (\psi ' , \theta ' ,
\phi ' )\}^{\ast} e^{-j(j+1) \gamma_{\rm so}^2 {\cal D} t}. 
\nonumber \\ 
\end{eqnarray}
Here an asterisk signifies a complex conjugate and 
Wigner's $D$ function is defined as
\beq
D_{m,n}^j (\psi , \theta , \phi ) = e^{ i m \phi} \, d_{m,n}^j (\theta ) \, 
e^{i n \psi } ,
\eeq
where
\beq
d_{m,n}^j (\theta ) &=&
\sqrt{ \frac{(j+m)! (j -m)!}{(j+n)! (j-n)!}}  \nonumber \\
&\times&
\cos^{m+n}(\theta /2) \sin^{m-n}(\theta /2) 
P_{j-m}^{(m-n,m+n)} (\cos\theta ) \nonumber \\
\eeq
in terms of the Jacobi polynomials $P_{k}^{(a,b)}(x)$. The index $j$ is an
integer or a half odd integer ($j=0,1/2,1,3/2,\cdots$ and
$m,n=-j,-j+1,\cdots ,j$). The conditional probability 
$g ( \omega ; t | \omega ' )$ satisfies a normalization condition
\beq
\int d \omega \, g ( \omega ; t | \omega ' ) 
=\int_{0}^{\pi}  \!\! d\theta  \int_{0}^{4\pi} \!\! d \phi \int_{0}^{4\pi} 
\!\! d \psi 
\,  \sin \theta \, g ( \omega ; t | \omega ' ) 
 = 1 . \nonumber 
\eeq

\section{Transport in the GOE-GSE crossover regime}

\subsection{Average conductance}

The average conductance $G$ is written in terms of the transmission 
amplitude $t_{a_1, a_2}$ as
\beq
\frac{G}{G_0}&=& \langle {\rm Tr} ({\bf t} {\bf t}^{\dagger} ) \rangle 
=\sum_{a_1 =1}^{N1}\sum_{a_2=1}^{N_2} {\rm Tr} \{ 
t_{a_1 , a_2} (t^{\dagger})_{a_2 , a_1} \} .
\eeq
Here the transmission matrix ${\bf t}$ is a $2 N_1 \times 2 N_2$ matrix which 
consists of the $2 \times 2$ blocks $t_{a_1,a_2}$. Then a semiclassical 
expression 
\beq
& & \langle {\rm Tr} ({\bf t} {\bf t}^{\dagger} ) \rangle 
\nonumber \\ &=& \frac{2}{T_H} 
\Bigl\langle \!
\sum_{a_1 , a_2} \sum_{\alpha, \gamma: a_1\to a_2} \!\!\!\!\!
{\cal A}_{\alpha} {\cal A}_{\gamma}^{\ast} 
\langle\!\langle {\rm Tr} (\Delta_{\alpha} \Delta_{\gamma}^{\dagger}) 
\rangle\!\rangle
e^{\frac{i}{\hbar} (S_{\alpha} - S_{\gamma})} \! 
\Bigr\rangle \nonumber \\   
\eeq
follows from (\ref{transmission}). Here the brackets $\langle ... 
\rangle$ mean an energy average, which eliminates the fluctuations 
of the physical quantities. If the difference between the actions 
$S_{\alpha}$ and $S_{\gamma}$ is sufficiently large, the exponential 
term $e^{\frac{i}{\hbar}(S_{\alpha} - S_{\gamma})}$ 
rapidly oscillates in the semiclassical limit $\hbar \rightarrow 0$, 
which eventually vanishes after averaging. Hence, in order to 
give a finite contribution, the trajectories  $\alpha$ and $\gamma$ 
are mutually almost the same. Then the identical trajectories 
$\alpha = \gamma$ yield the first order approximation, which is referred 
to as ``the diagonal approximation''\cite{berry}. These mutually 
identical trajectory pairs yield the following contribution 
\beq
\langle {\rm Tr} ({\bf t} {\bf t}^{\dagger} ) \rangle_1 
&=& \frac{2}{T_H} \sum_{a_1 , a_2 } \sum_{\alpha } 
{\rm Tr} (\Delta_{\alpha} \Delta_{\alpha}^{\dagger} )
| {\cal A}_{\alpha } |^{2}
 \nonumber \\ 
&=& \frac{4}{T_H} \sum_{a_1, a_2 } \sum_{\alpha } | {\cal A}_{\alpha } |^{2}
\nonumber \\
&=& \frac{4}{T_H} N_1 N_2 \int_{0}^{\infty} dT e^{-(2N/T_H ) T}  \nonumber \\ 
&=& \frac{2 N_1 N_2}{N} ~ .
\eeq
Here we used the sum rule (\ref{A_decay}) for the 
stability amplitude ${\cal A}_{\alpha}$. In this calculation, 
a product of the spin matrices is reduced to the identity matrix, 
and the trace yields a factor $2$. The diagonal approximation does not 
discriminate the symmetry classes.
\par
The second order approximation comes from the 
Richter-Sieber (RS) pairs \cite{rs02,sr_pair} drawn 
in Fig.1. In the RS pair, two trajectories 
come close to each other in the encounter region, and go 
in the opposite directions on one loop. We can symbolically 
write RS pairs (See Fig.1) as 
\beq
\alpha &:& L_1 E L_2 \bar{E} L_3  ~ , \nonumber \\
\gamma &:& L_1 E \bar{L}_2 \bar{E} L_3 ~,\nonumber
\eeq
where $E$ implies one of the two trajectory segments in 
the encounter region where two loops are connected and $\overline{E}$ implies 
the time reverse of $E$. The loops are denoted as $L_1$, $L_2$ and $L_3$, 
respectively, and $\overline{L}_j$ $(j=1,2,3)$ is the time reverse of $L_j$. 
Using these notations, we can write the RS pair contribution as
\beq
& & \langle {\rm Tr} ({\bf t} {\bf t}^{\dagger} ) \rangle_2  \nonumber \\ 
& = & \frac{2}{T_H} \sum_{a_1 , a_2 } 
\sum_{\substack{\alpha: L_1 E L_2 \bar{E} L_3 \\ \gamma: 
L_1 E \bar{L}_2 \bar{E} L_3}}
\!\!\!\!\!\! \Bigl\langle
{\cal A}_{\alpha} {\cal A}_{\gamma}^{\ast} 
\langle\!\langle {\rm Tr}(\, \Delta_{\alpha}
\Delta_{\gamma}^{\dagger}  \, )
 \rangle\!\rangle e^{\frac{i}{\hbar} \Delta S} \Bigr\rangle, 
\nonumber \\ 
\label{g2}
\eeq
where $\Delta S$ is the action difference $S_{\alpha} - S_{\gamma}$. 
The spin matrices $\Delta_{\alpha}$ and $\Delta_{\gamma}$ are factored 
into the loop and encounter parts as 
\beq
{\Delta_{\alpha}} &=&
\Delta_{L_1} \Delta_{E}\Delta_{L_2} (\Delta_{E})^{-1} \Delta_{L_3} ~ ,\\
\Delta_{\gamma} &=&
\Delta_{L_1} \Delta_{E} (\Delta_{L_2})^{-1}(\Delta_{E})^{-1}
\Delta_{L_3}. 
\eeq
Along the trajectories the non-Abelian nature of the spin operators 
must be taken into account. A time reversal operation of a spin matrix 
is realized by a matrix inversion. 

\begin{figure}[h]
\epsfxsize=8cm
\centerline{\epsfbox{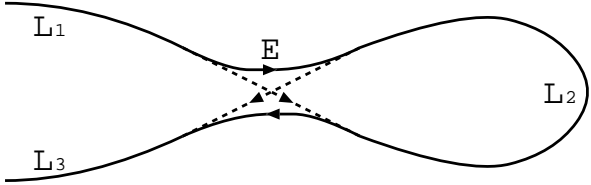}}
\caption{The Richter-Sieber (RS) pair. 
The solid and dashed curves are respectively 
$\alpha$ and $\gamma$ orbits in the text.}
\end{figure}

We divide the whole time elapsed on a trajectory into the loop and 
encounter parts, i.e., $T_1, T_2, T_3$ for $L_1$, $L_2$, $L_3$, respectively, 
and $t_{\rm enc}$ for $E$. It should be noted that the existence 
of the encounter affects the survival probability $e^{-(2 N/T_H)T}$ 
in Eq.~(\ref{A_decay})\cite{braun,haake}: 
if one of the two orbit segments 
in the encounter is inside the cavity, the other segment must also 
remain inside. Hence the survival probability is modified 
into $e^{-(2 N/T_H) (T-t_{\rm enc})}$. 
\par
In the encounter region, the classical 
actions of the two trajectories are slightly different. 
The action difference can be estimated by using the coordinates 
$(s,u)$ along the stable and unstable manifolds within the ranges 
$s,u \in [-c,c]$\cite{braun,haake}. The time duration $t_{\rm enc}$ 
inside the encounter region is related to the Lyapunov exponent 
$\lambda$ as $t_{\rm enc} \sim \frac{1}{\lambda} \ln \frac{c^{2}}{|us|}$, 
and the action difference is expressed as 
\beq
\Delta S = us.
\eeq
The number density of encounters in a trajectory with an elapsed time 
$T=T_1+T_2+T_3 +2 t_{\rm enc}$ is evaluated as\cite{braun,haake}  
\beq
\omega (s,u) ds du = 
\int_{\substack{T_1,T_2 >0 \\ T_1 + T_2 < T - 2 t_{\rm enc}}}
d T_1 dT_2 \frac{1}{t_{\rm enc} \Omega /2} ds du
\eeq
by taking account of $T_1, T_2 > 0 $ and $T_3= T - T_1 -T_2 - 2 
t_{\rm enc} > 0$.
\par
On the other hand, the spin diffusion term is calculated as
\beq
\langle\!\langle {\rm Tr}(\Delta_{\alpha}
\Delta_{\gamma}^{\dagger}  )
\rangle\!\rangle 
&=&\int d \omega_{L_1} d \omega_{L_2}d \omega_{L_3} d \omega_{E_1}
{\rm Tr}
\left\{ (\Delta_{L_2})^2 \right\} \nonumber \\
&\times& g(\omega_{L_1} , T_1 | 0,0,0)
g(\omega_{L_2} , T_2 | 0,0,0) \nonumber \\
&\times&
g(\omega_{L_3} , T_3 | 0,0,0)
g(\omega_{L_E} , t_{\rm enc} | 0,0,0) 
\nonumber \\
&=& -1 + 3 e^{- 2 \gamma_{\rm so}^2 {\cal D} T_2}. \label{instance} 
\eeq
Using the above formulas, we find that 
the RS contribution is
\beq
\langle {\rm Tr} ({\bf t} {\bf t}^{\dagger} ) \rangle_2  &=& 
\frac{2 N_1 N_2}{T_H}
\int_{-c}^{c} ds du
\int_{t_{\rm enc}}^{\infty} dT 
 \nonumber \\
&\times &\omega (s,u) \, 
e^{-(2N/T_H)(T-t_{\rm enc})}
\langle\!\langle {\rm Tr}(\Delta_{\alpha}
\Delta_{\gamma}^{\dagger}  )
\rangle\!\rangle 
\nonumber \\
&=&
\frac{2 N_1 N_2}{T_H}
\int_{0}^{\infty} d T_1 d T_2 d T_3 \int_{-c}^{c} ds du 
~ \frac{2 e^{i su/\hbar}}{\Omega t_{\rm enc}} 
\nonumber \\
&\times&
e^{-(2N/T_H)(T_1 + T_2  + T_3 + t_{\rm enc})} 
\, (3 e^{-2\gamma_{\rm so}^2 {\cal D} T_2} -1) 
\nonumber \\
& = & \frac{N_1 N_2}{(N_1 + N_2 )^2} \left( 1 - 
\frac{3}{1 + \gamma_{\rm so}^2 {\cal D} T_H/N }\right).
\eeq
The last line of the above formula is obtained by the 
following criterion: after expanding the formula in $t_{\rm enc}$ 
and integrating each term over $(s,u)$, any term dependent on $t_{\rm enc}$ 
vanishes in the semiclassical limit, and a finite contribution 
comes only from the terms independent of $t_{\rm enc}$.
\par
The third order contribution to the average conductance comes 
from the diagrams drawn in Fig.2. In Appendix A, the spin diffusion 
terms are listed. Using these results, we arrive at an 
expression of the average conductance 
\beq
\frac{G}{G_0} & = & \frac{2 N_1 N_2}{N} + 
\frac{N_1 N_2}{N^2} \left( 1 - \frac{3}{1 + \xi} \right)
\nonumber \\  
& + &  \frac{N_1 N_2}{2 N^3} 
\left\{ 1  - \frac{3}{1 + \xi} 
+ \frac{3}{(1 + \xi )^2} 
+ \frac{3}{(1 + \xi )^3} 
\right\} \nonumber \\ 
& + & O(1/N^2), \label{result_g}
\eeq
where $\xi$ is the crossover parameter defined as $\xi=
\gamma_{\rm so}^2 {\cal D} T_H/N $.
We can easily check that (\ref{result_g}) reproduces 
the $1/N$ expansions of the GOE and GSE formulas 
(\ref{rdm_conductance}) by taking the limits 
$\xi\to 0$ and $\xi\to\infty$, respectively. It follows 
from this result that only one parameter $\xi$ 
is necessary to describe the GOE-GSE crossover. 

\begin{widetext}

\begin{figure}[h]
\epsfxsize=16cm
\centerline{\epsfbox{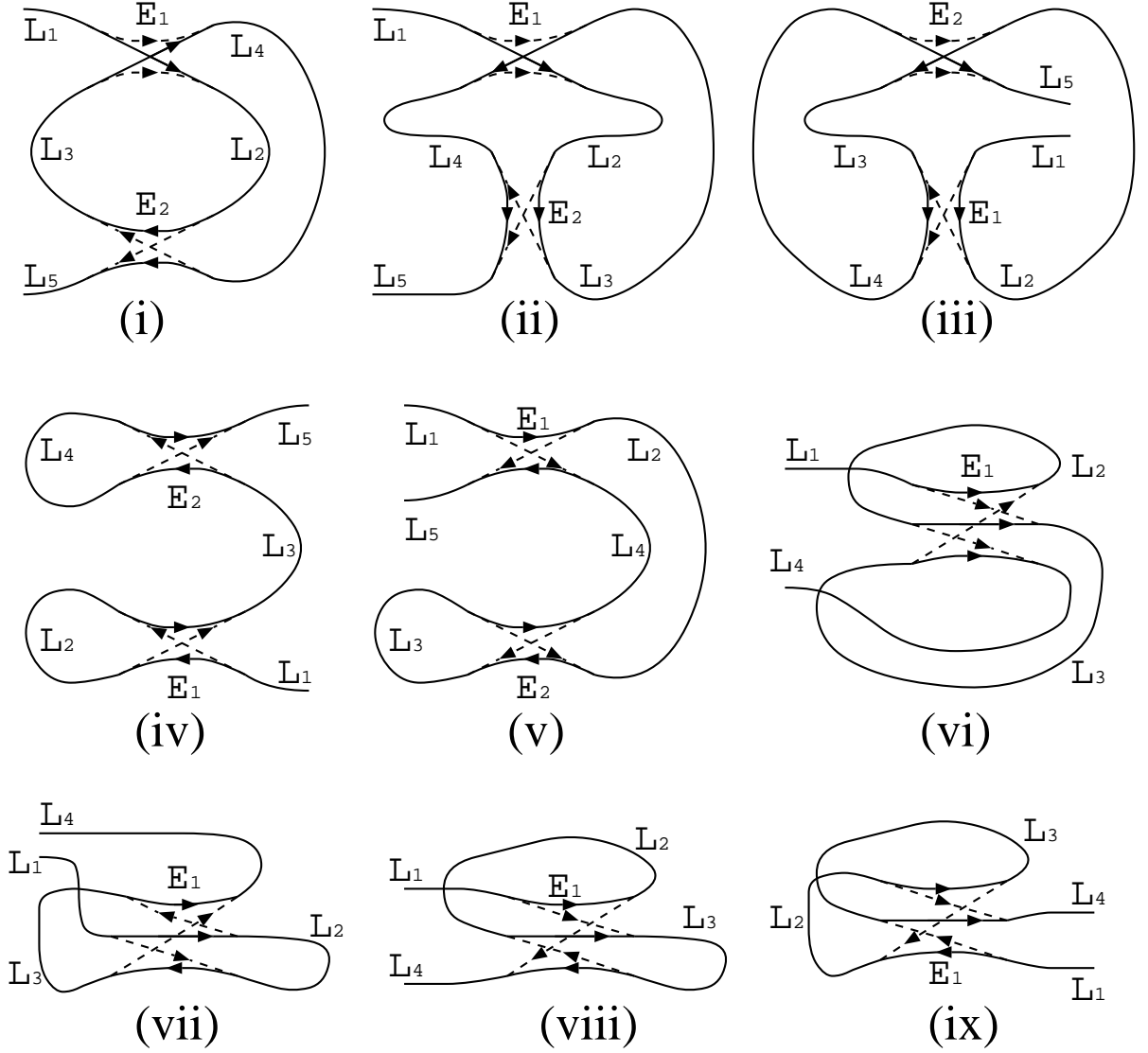}}
\caption{The diagrams of the third order contributing to the conductance}
\end{figure}

\end{widetext}

\subsection{Conductance variance}

The conductance variance ${\rm Var} G$ is written in terms of the 
transmission matrix ${\bf t}$ as
\beq
\frac{{\rm Var} G}{G_0^2} &=& \left\langle 
\left\{ {\rm Tr}( {\bf t} {\bf t}^{\dagger}) \right\}^2 \right\rangle
- \langle{\rm Tr} ({\bf t} {\bf t}^{\dagger})\rangle^2.
\eeq
The semiclassical expression of the first term is
\beq
\left\langle \left\{ {\rm Tr} ( t t^{\dagger}) \right\}^2 
\right\rangle & = &
\frac{4}{T_H^2}
\sum_{a_1 , a_2}\sum_{c_1 , c_2  }
\sum_{\alpha, \beta : a_1 \to a_2 }\sum_{\gamma , \delta : c_1 \to c_2 }
\nonumber \\
&\times&\Bigl\langle{\cal A}_{\alpha } {\cal A}_{\beta }^{\ast}
{\cal A}_{\gamma } {\cal A}_{\delta }^{\ast}
e^{i ( S_{\alpha} - S_{\beta} + S_{\gamma } - S_{\delta } )/\hbar} 
\nonumber \\
& \times &
\langle\!\langle 
{\rm Tr } (\Delta_{\alpha }\Delta_{\beta }^{\dagger} )
{\rm Tr } (\Delta_{\gamma }\Delta_{\delta }^{\dagger} )
\rangle\!\rangle  \Bigr\rangle ~.
\eeq
Let us first adopt the diagonal approximation, for which 
we have two types of contributions. One is given by setting 
$\alpha=\beta$ and $\gamma=\delta$, where $a_1,a_2,c_1$ and 
$c_2$ are all independent. The other choice is $\alpha=\delta, 
\gamma=\beta$, and $a_1=c_1,\, a_2=c_2$.  The numbers of possible 
channel combinations are $N_1^2 N_2^2 $ and $N_1 N_2$, respectively. 
These contributions are summed up to yield 
\beq
&& \left\langle \left\{ {\rm Tr} ({\bf t} {\bf t}^{\dagger}) 
\right\}^2 \right\rangle_1
= \frac{4}{T_H^2}
\Bigl \langle 
4\sum_{\substack{a_1, c_1 \\ a_2, c_2}} 
\sum_{\substack{\alpha=\beta : a_1 \to a_2 \\ 
\gamma=\delta : c_1 \to c_2}} |{\cal A}_{\alpha }|^2 |{\cal A}_{\gamma }|^2
\nonumber \\
&+&
\sum_{\substack{a_1= c_1 \\ a_2 = c_2}} 
\sum_{\substack{\alpha=\delta : a_1 \to a_2 \\ 
\gamma=\beta : a_1 \to a_2}} |{\cal A}_{\alpha }|^2 |{\cal A}_{\gamma }|^2
\langle\!\langle 
|{\rm Tr}(\Delta_{\alpha} \Delta_{\gamma}^{\dagger } )|^2
\rangle\!\rangle
\Bigr\rangle.~~~~~
\eeq
Here the spin diffusion term is 
\beq
& & \langle\!\langle 
|{\rm Tr}(\Delta_{\alpha} \Delta_{\gamma}^{\dagger } )|^2
\rangle\!\rangle \nonumber \\ 
&=& \int d \omega_{\alpha} d \omega_{\gamma}  
g(\omega_{\alpha} , T_{\alpha} | 0,0,0) g(\omega_{\gamma} , T_{\gamma} | 0,0,0) 
\nonumber \\
&=&
1 +3 e^{-2 \gamma_{\rm so}^2 {\cal D} (T_{\alpha} + T_{\gamma})},
\eeq
where $T_{\alpha}$ and $T_{\gamma}$ are the times elapsed on 
the trajectories $\alpha$ and $\gamma$, respectively.  
Using this expression, one can obtain the diagonal contribution
\beq
\left\langle \left\{ {\rm Tr} ( {\bf t} {\bf t}^{\dagger}) 
\right\}^2 \right\rangle_1 &=&  
\frac{4 N_1^2 N_2^2}{N^2}+
\frac{N_1 N_2}{N^2} \left\{ 1 + \frac{3}{(1 + \xi )^2} \right\}. ~~~~~~~
\eeq
To go beyond the diagram approximation, we note that the next order 
diagrams are classified into $d$-families and $x$-families as shown 
in Fig.3\cite{haake}: $d$-quadruplets are drawn in the diagrams (i)-(vii), 
while $x$-quadruplets in (viii). 

\begin{widetext}

\begin{figure}[h]
\epsfxsize=16cm
\centerline{\epsfbox{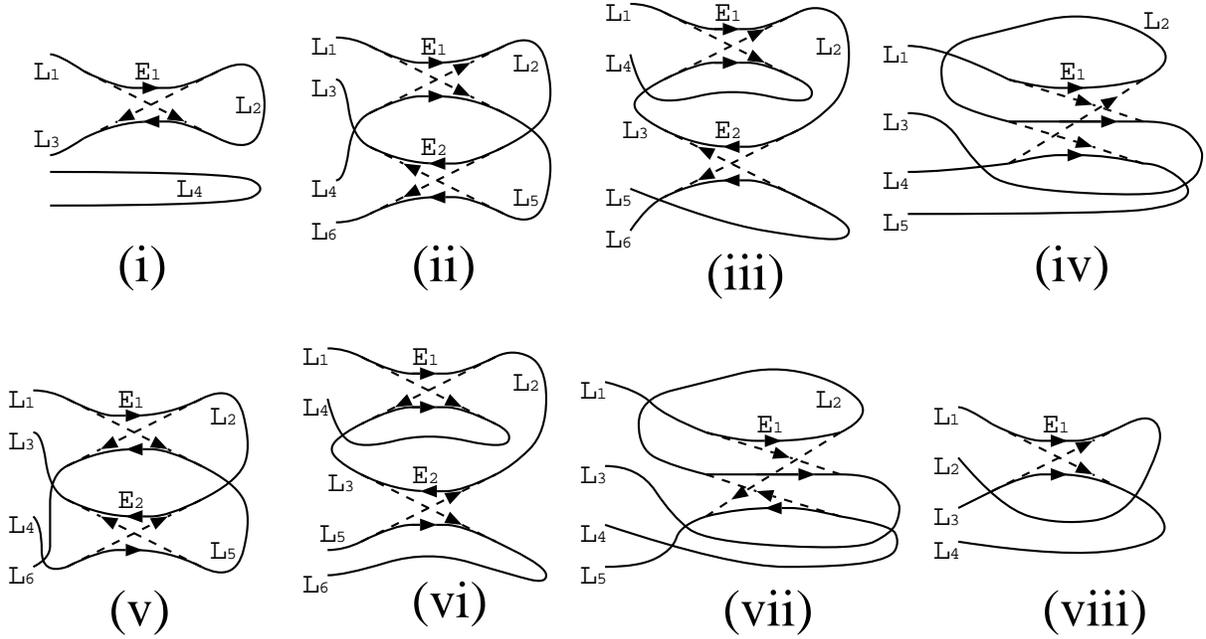}}
\caption{The diagrams contributing to the conductance variance} 
\end{figure}

\end{widetext}

Let us write the next order term as
\beq
& & \left\langle \left\{ {\rm Tr} ({\bf t} {\bf t}^{\dagger}) 
\right\}^2 \right\rangle_2 \nonumber \\ 
& = & (N_1^2 N_2^2 + N_1 N_2 ) \left( \frac{d_1}{N^3} 
+ \frac{d_2}{N^4} \right) + N_1 N_2 \frac{x_1}{N^2}. \nonumber
\eeq 
Here the coefficients $d_1$, $d_2$ and $x_1$ are obtained from the 
families of quadruplets. The coefficient $d_1$ comes from the 
diagram (i): quadruplets consisting of one diagonal 
pair and one RS pair. On the other hand, many 
diagrams have to be taken into account to calculate $d_2$, i.e., 
$1$): quadruplets consisting of one diagonal pair and one pair 
contributing to the $O(1/N)$ term in the expansion (\ref{result_g}) 
of the average conductance, $2$): two RS pairs $3$): the diagrams (ii)-(vii) 
shown in Fig.3. The coefficient $x_1$ is calculated from the 
diagram (viii) in Fig.3.  Considering the contributions from 
these diagrams, we obtain 
\beq
d_1 & = & 4 - \frac{12}{1 + \xi}, \nonumber \\ 
d_2 & = & 5 - \frac{12}{1 + \xi} + 
\frac{21}{(1 + \xi)^2} + \frac{6}{(1 + \xi)^3}, \nonumber \\ 
x_1 & = & - 1 - \frac{3}{(1 + \xi)^2}. \nonumber
\eeq
Then we find the expression of the conductance variance for 
$N_1 \gg 1$ and $N_2\gg 1$ as 
\beq
\frac{{\rm Var}G}{G_0^2}&=& \frac{N_1^2 N_2^2}{N^4}
\left\{ 2 + \frac{6}{(1 + \xi )^2}\right\} + O\left( \frac{1}{N} \right). 
\label{result_varg}
\eeq
One can check that the GOE and GSE limits agree with 
(\ref{rdm_conductance_fluc}). 

\subsection{Shot noise}

The shot noise $P$ is written in the form 
\beq
\frac{P}{P_0} &=& 
\langle {\rm Tr} ({\bf t} {\bf t}^{\dagger} - {\bf t} 
{\bf t}^{\dagger} {\bf t} {\bf t}^{\dagger} )\rangle .
\eeq
The second term is semiclassically expressed as 
\beq
\langle {\rm Tr} ({\bf t} {\bf t}^{\dagger} {\bf t} {\bf t}^{\dagger} ) \rangle
&=& \frac{4}{T_H^2} 
\sum_{\substack{{a_1, a_2 } \\  {c_1, c_2}}}
\sum_{
{\substack{\alpha : a_1 \to a_2,
\beta : c_1 \to a_2  \\ 
\gamma : c_1 \to c_2,
\delta : a_1 \to c_2 }}
} \nonumber \\
&\times&
\Bigl\langle A_{\alpha}A_{\beta}^{\ast}A_{\gamma}A_{\delta}^{\ast}
e^{\frac{i}{\hbar} (S_{\alpha} - S_{\beta} + S_{\gamma} - S_{\delta })} 
\nonumber \\
&\times&
\langle\!\langle 
{\rm Tr} ( \Delta_{\alpha}\Delta_{\beta}^{\dagger}
\Delta_{\gamma}\Delta_{\delta}^{\dagger})
\rangle\!\rangle \Bigr\rangle. 
\eeq
The diagonal contribution consists of two terms. One term has 
$\alpha=\beta$ and $\gamma=\delta$ where we need to set $a_1=c_1$.
The other term has $\alpha=\delta , \, \beta=\gamma$, and $a_2=c_2$.
We sum up these two terms and obtain  
\beq
&&\langle {\rm Tr}({\bf t} {\bf t}^{\dagger} 
{\bf t} {\bf t}^{\dagger}) \rangle_1  \nonumber \\
&=& \frac{16}{T_H^2} \Bigl\langle 
\sum_{a_1, a_2, c_2 }
\sum_{\substack{\alpha : a_1 \to a_2 \\
\gamma : a_1 \to c_2 }} |A_{\alpha}|^2|A_{\gamma}|^2
\langle\!\langle 
{\rm Tr} ( \Delta_{\alpha}\Delta_{\alpha}^{\dagger}
\Delta_{\gamma}\Delta_{\gamma}^{\dagger})
\rangle\!\rangle  \nonumber \\
&&~~~+
\sum_{a_1, a_2, c_1 }
\sum_{\substack{\alpha : a_1 \to a_2 \\ 
\beta : c_1 \to a_2 }} |A_{\alpha}|^2|A_{\beta}|^2
\langle\!\langle 
{\rm Tr} ( \Delta_{\alpha}\Delta_{\beta}^{\dagger}
\Delta_{\beta}\Delta_{\alpha}^{\dagger})
\rangle\!\rangle  
\Bigr\rangle 
 \nonumber \\
&=& 2 \frac{N_1 N_2}{N} .
\eeq
It follows that the diagonal contributions to 
$\langle {\rm Tr} ({\bf t} {\bf t}^{\dagger})\rangle$ and 
$\langle {\rm Tr} ({\bf t} {\bf t}^{\dagger} {\bf t} {\bf t}^{\dagger})\rangle $ 
are both $2 \frac{N_1 N_2}{N}$, and mutually cancel.
\par
Hence the RS pair contribution to ${\rm Tr}({\bf t} {\bf t}^{\dagger})$, 
and the contribution to ${\rm Tr}({\bf t} {\bf t}^{\dagger} {\bf t} {\bf t}^{\dagger})$ from the quadruplets drawn in Fig.3 (i) and (viii) have to be 
calculated. Moreover we take account of the additional diagrams 
shown in Fig.4\cite{braun,haake}. Summing up these contributions, 
we finally obtain the shot noise in the crossover regime as
\beq
\frac{P}{P_0}
&=& 2 \frac{N_1^2 N_2^2}{N^3} + \frac{N_1 N_2 (N_1 - N_2)^2}{N^4}
\left( \frac{3}{1 + \xi}-1  \right) \nonumber \\ 
& + & O(1/N). \label{result_P}
\eeq
Let us denote the $O(1)$ terms of $G$ and $P$ by $\delta G$ and $\delta P$, 
respectively. It can be seen from Eqs.~(\ref{result_g}) and (\ref{result_P}) 
that 
\beq
\frac{\delta P/P_0}{\delta G/G_0} = 
- \left( \frac{N_1 - N_2}{N_1 + N_2} \right)^2,
\eeq
which is a universal relation established in Ref.\onlinecite{beri2}.     

\begin{widetext}

\begin{figure}[h]
\epsfxsize=16cm
\centerline{\epsfbox{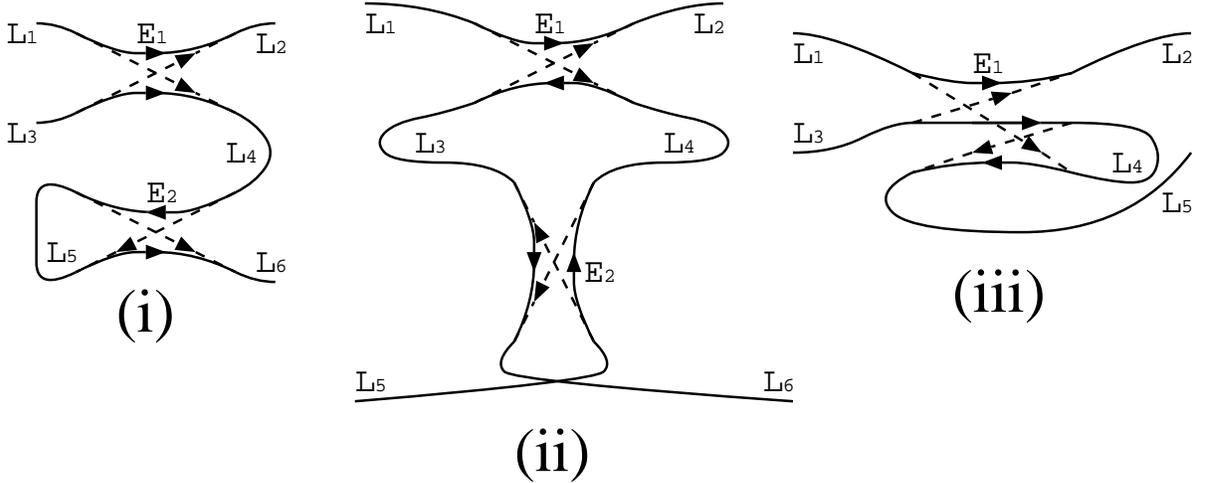}}
\caption{The diagrams contributing to the shot noise}
\end{figure}

\end{widetext}

\section{Comparison with a random matrix model}

In this section, a random matrix model on the GOE-GSE 
crossover is numerically analyzed and the results are 
compared with the semiclassical formulas. In the theory 
of random matrices, a time-reversal invariant quantum Hamiltonian 
with spin $1/2$ is simulated by a self-dual real quaternion 
random matrix\cite{mehta}. A real quaternion $q$ is a linear combination
\begin{equation}
q = q_0 e_0 + q_1 e_1 + q_2 e_2 + q_3 e_3,
\end{equation}
where $q_j$ are real numbers and called the $j$-th component 
of $q$. The bases $e_0,e_1,e_2,e_3$ can be represented 
by $2 \times 2$ matrices as
\begin{eqnarray}
\label{2times2}
& & e_0 \rightarrow \left( \begin{array}{cc} 1 & 0 \\ 0 & 1 \end{array} 
\right), \ \ \  
e_1 \rightarrow \left( \begin{array}{cc} i & 0 \\ 0 & -i \end{array} 
\right), \nonumber \\    
& & e_2 \rightarrow \left( \begin{array}{cc} 0 & 1 \\ -1 & 0 \end{array} 
\right), \ \ \  
e_3 \rightarrow \left( \begin{array}{cc} 0 & i \\ i & 0 \end{array} 
\right),
\end{eqnarray}   
so that $q_0 e_0$ is equated with a real number $q_0$. The dual 
quaternion of $q$ is defined as
\begin{equation}
{\bar q} = q_0 e_0 - q_1 e_1 - q_2 e_2 - q_3 e_3.
\end{equation}
When an $m \times n$ real quaternion matrix $Q$ has the 
$(j,l)$ element $q_{jl}$, we define that the $n \times m$ 
dual matrix ${\bar Q}$ has the $(j,l)$ element ${\bar q}_{lj}$. 
If a square real quaternion matrix satisfies $Q = {\bar Q}$, 
then $Q$ is called a self-dual real quaternion matrix.      
\par
The parametric motion of a self-dual real quaternion 
random matrix is realized in the framework of Dyson's 
matrix Brownian motion model\cite{dyson}. It is postulated that 
the probability density function (p.d.f.) of an 
$M \times M$ self-dual real quaternion matrix $H$ is  
\begin{equation}
\label{PHR}
P(H;\tau | R) \ dH \propto
{\rm exp}\left\{ - 2 \frac{{\rm Tr}(H - e^{-\tau} R)^2}
{1 - e^{- 2 \tau}} \right\} dH
\end{equation}
with
\begin{equation}
dH = \prod_{j=1}^M dH_{jj} \prod_{j<l}^M
\prod_{k=0}^3 dH^{(k)}_{jl}.
\end{equation}
Here $H^{(k)}_{jl}$ is the $k$-th component of $H_{jl}$. 
We are interested in the parametric motion of the matrix 
$H$ depending on the fictitious time parameter $\tau$. 
\par
At the initial time $\tau = 0$, this p.d.f. is 
reduced to
\begin{equation}
P(H;0| R) \ dH = \delta(H - R) \ dH,
\end{equation}
so that the self-dual real quaternion matrix $R$ gives the 
initial condition of the parametric motion. On the other hand, 
in the limit $\tau \rightarrow \infty$,      
\begin{equation}
P(H;\infty | R) \ dH \propto
{\rm exp}\left( - 2 {\rm Tr}H^2 \right) dH,
\end{equation}
which is the p.d.f. of the GSE.
\par
Let us suppose that the elements of the initial matrix $R$ 
have only the $0$-th components ($R$ is then a real symmetric 
matrix) and that the p.d.f. of $R$ is 
\begin{equation}
P_{\rm GOE}(R) dR \propto {\rm exp}\left( - \frac{1}{2} 
{\rm Tr} R^2  \right) dR
\end{equation}
with
\begin{equation} 
dR = \prod_{j=1}^M dR_{jj} \prod_{j<l}^M dR_{jl},
\end{equation}
which is the p.d.f. of the GOE. Then we can calculate the p.d.f. of $H$ 
at a fictitious time 
$\tau$ as 
\begin{eqnarray}
\label{crossover}
& & P(H) dH = 
\left\{ \int dR \ P(H;\tau | R) P_{\rm GOE}(R) \right\} dH 
\nonumber \\ 
& \propto &  
\left[ \int dR \  
{\rm exp}\left\{ - 2 \frac{{\rm Tr}(H - e^{- \tau} R)^2}
{1 - e^{- 2 \tau}} - \frac{1}{2} {\rm Tr} R^2 \right\} \right]  
dH  \nonumber \\ 
& \propto & 
{\rm exp}\left[ 
- \frac{2}{1 + 3 e^{- 2 \tau}} \sum_{j=1}^M (H^{(0)}_{jj})^2 
- \frac{4}{1 + 3 e^{- 2 \tau}} \sum_{j<l}^M (H^{(0)}_{jl})^2  
\right. \nonumber \\ 
& & \left. - \frac{4}{1 - e^{- 2 \tau}} \sum_{j<l}^M 
\left\{ (H^{(1)}_{jl})^2 + (H^{(2)}_{jl})^2 + (H^{(3)}_{jl})^2 
\right\} \right] dH, \nonumber \\ 
\end{eqnarray} 
which describes the crossover from the GOE (at $\tau = 0$) to 
GSE (at $\tau = \infty$). The components of the elements $H_{jl}$ ($j \leq l$) are 
independently distributed according to Gaussian density functions. 
\par
If an $M \times M$ real quaternion matrix $Z$ satisfies
\begin{equation}
Z {\bar Z} = {\bar Z} Z = I_M
\end{equation} 
($I_M$ is the $M \times M$ identity matrix), then $Z$ is called 
a symplectic matrix. If the elements of a symplectic matrix $U$ 
only have the $0$-th elements, then $U$ is a real orthogonal matrix. 
It is known that the measure $dH$ is invariant 
under the symplectic transformation $H \mapsto {\bar Z} H Z$ 
and the measure $dS$ is invariant under the orthogonal 
transformation $S \mapsto U^{\rm T} S U$ ($U^{\rm T}$ is 
the transpose of $U$). It follows that the p.d.f. $P(H) 
dH$ in (\ref{crossover}) is invariant under the 
orthogonal transformation $H \mapsto U^{\rm T} H U$.
\par
We go back to the problem of a chaotic cavity with $M$ 
bound states to which two leads with $N_1$ and $N_2$ 
propagating modes are attached($M \geq N = N_1 + N_2$). 
We are interested in the limit $M \rightarrow \infty$. 
Let us suppose that the $M \times M$ matrix $H$ describing 
the scattering in the cavity is a random matrix distributed 
according to the crossover p.d.f. in (\ref{crossover}). Then 
the $N \times N$ scattering matrix $S$ is\cite{beenakker97} 
\begin{equation}
S = I_N + i 2 \pi W^{\rm T} (H-E_F-i \pi W W^{\rm T} )^{-1} W,
\end{equation}
where $E_F$ is the Fermi energy, and the elements of an $M \times N$ 
real matrix $W$ are the coupling constants between the cavity 
and the leads.
\par
Assuming that the tunnel probability of the leads 
are $1$, we can see that the eigenvalues of 
$W^{\rm T} W$ are all $M/(\rho \pi^2)$, where
$\rho$ is the eigenvalue density of $H$ at 
the Fermi energy. Then a singular value 
decomposition
\begin{equation}
W = U D V 
\end{equation}      
holds, where $U$ and $V$ are $M \times M$ 
and $N \times N$ real orthogonal matrices, respectively, 
and $D$ is an $M \times N$ matrix
\begin{equation}
D = \frac{1}{\pi} \sqrt{\frac{M}{\rho}} {\tilde W} 
\end{equation}
with
\begin{equation}
{\tilde W} = \left( \begin{array}{c} I_N \\ O \end{array} \right).
\end{equation}   
Here $O$ is an $(M-N) \times N$  matrix consisting 
of zero elements. When the Fermi energy 
$E_F$ is set to zero (so that $\rho = \sqrt{2 M}/\pi$), 
the scattering matrix $S$ can be rewritten as
\begin{equation}
\label{Smatrix}
S = I_N + i \sqrt{2 M} {\tilde W}^{\rm T} \left( 
{\tilde H} - i \sqrt{\frac{M}{2}} {\tilde W} {\tilde W}^{\rm T} \right)^{-1} 
{\tilde W},
\end{equation}
where
\begin{equation}
{\tilde H} = (U {\tilde V})^{\rm T} H (U {\tilde V})
\end{equation}
with
\begin{equation}
{\tilde V} = \left( \begin{array}{cc} V & O \\ O & I_{M-N} 
\end{array} \right).
\end{equation}    
Since $U {\tilde V}$ is a real orthogonal matrix, the 
$M \times M$ matrix ${\tilde H}$ is also distributed 
according to the p.d.f. in (\ref{crossover}) 
(with $H$ replaced by ${\tilde H}$).  
\par
Thus the scattering matrix $S$ can numerically be 
generated by using the Gaussian p.d.f. (\ref{crossover}) 
of ${\tilde H}$ and the relation (\ref{Smatrix}). Replacing 
the quaternion elements of $S$ by the $2 \times 2$ matrix 
representations (\ref{2times2}), we obtain a $2 N \times 2 N$ 
matrix ${\tilde S}$. It is written in terms of the $2 N_1 
\times 2 N_2$ transmission matrix ${\bf t}$ as 
\begin{equation}
{\tilde S} = \left( \begin{array}{cc} {\bf r} & {\bf t}^{\dagger} \\ 
{\bf t} & {\bf r'} \end{array} \right), 
\end{equation}
where ${\bf r}$ and ${\bf r'}$ are the reflection matrices. Then the average 
conductance $G$ can be evaluated as
\begin{equation}
\frac{G}{G_0} 
= \langle {\rm Tr} ({\bf t} {\bf t}^{\dagger}) \rangle = \left\langle 
\sum_{j=2 N_2+1}^{2 (N_1 + N_2)} \sum_{l=1}^{2 N_2} 
|{\tilde S}_{jl}|^2 \right\rangle, 
\end{equation}  
where the brackets $\langle ... \rangle$ denote an average 
over the p.d.f. (\ref{crossover}). The conductance variance 
${\rm Var}G$ and the shot noise $P$ 
can similarly be written as
\begin{equation}
\frac{{\rm Var}G}{G_0^2} = 
\left\langle \left\{ {\rm Tr}({\bf t} {\bf t}^{\dagger}) 
\right\}^2 \right\rangle - 
\langle {\rm Tr}({\bf t} {\bf t}^{\dagger}) \rangle^2
\end{equation} 
and
\begin{equation}
\frac{P}{P_0} 
=  \left\langle {\rm Tr}({\bf t} {\bf t}^{\dagger}) 
- {\rm Tr}({\bf t} {\bf t}^{\dagger} {\bf t} {\bf t}^{\dagger}) \right\rangle.
\end{equation} 
\par
The remaining task is to find a relation between the semiclassical 
parameter $\xi$ and the fictitious time $\tau$. Pandey analyzed hierarchical 
relations among the eigenvalue correlation functions of random matrices 
and evaluated the form factor $K(k;\tau)$ (the Fourier transform of 
the scaled two eigenvalue correlation function). For the random matrices 
obeying the crossover p.d.f. in (\ref{crossover}), he derived 
a relation\cite{pandey}
\begin{equation}     
K(k;\tau) = K(k;\infty) + \left\{ K(k;0) - K(k;\infty) \right\} 
e^{- 4 \pi^2 \rho^2 \tau k} 
\end{equation}
with $k \downarrow 0$. Here $K(k;0)$ and $K(k;\infty)$ are the form 
factors of the GOE and GSE random matrices, respectively. They are known 
to be\cite{mehta}
\begin{equation}
K(k;0) = 2 k, \ \ \ K(k;\infty) = \frac{k}{2}, \ \ \ k \downarrow 0, 
\end{equation}    
so that 
\begin{equation}
\label{ffrm}     
K(k;\tau) = \frac{k}{2} \left( 1 + 3 \  
e^{- 8 M \tau k} \right), \ \ \ k \downarrow 0. 
\end{equation}
On the other hand, Nagao and Saito semiclassically analyzed the 
form factor of a chaotic system with a weak spin-orbit interaction. 
They obtained a small $k > 0$ expansion up to the second 
order\cite{nagao_saito} 
\begin{equation}
\label{ffsc}
K(k;a) = \frac{k}{2} ( 1 + 3 \ e^{- a k T_H}) + 
\frac{k^2}{4} \left\{ 1 + (3 a k T_H - 9) \ e^{- a k T_H} \right\}
\end{equation}
with $a = \gamma_{\rm so}^2 {\cal D}$. Comparing (\ref{ffrm}) and 
(\ref{ffsc}), we arrive at a relation
\begin{equation}
8 M \tau = a T_H.
\end{equation}
Therefore the semiclassical parameter $\xi = a T_H/N$ is associated 
with the random matrix parameter $\tau$ as  
\begin{equation}
\xi = \frac{8 M}{N} \tau.
\end{equation}

\begin{figure}[h]
\epsfxsize=9cm
\centerline{\epsfbox{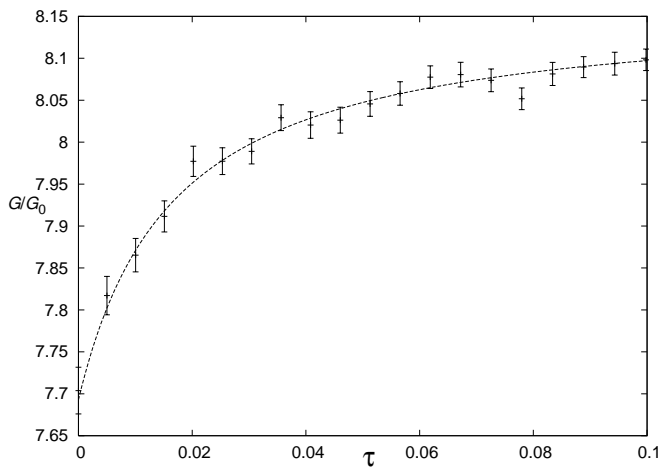}}
\caption{A comparison of the semiclassical result (curve) and 
numerically calculated random matrix results (errorbars) for 
$G/G_0$.}
\end{figure}

\par
In Fig.5, numerical calculations of $G/G_0$ at various values 
of $\tau$ are compared with the corresponding semiclassical predictions 
(\ref{result_g}) with $\xi = 8 M \tau/N$. In the numerical 
calculations, we set 
$M = 200$, $N_1 = 20$ and $N_2 = 5$. The errorbars are introduced 
in order to estimate the statistical errors due to the fact that 
the averages are calculated over only $300$ samples. Note that 
the semiclassical formulas are truncated and 
hence are valid only for large $N_1$ and $N_2$. Nevertheless we 
can see a fairly reasonable agreement with the numerical results. 
\par
Similar plots for ${\rm Var}G/G_0^2$ and $P/P_0$ are also shown, 
respectively, in Figs. 6 and 7. The semiclassical curves are 
drawn by using eqs. (\ref{result_varg}) and (\ref{result_P}). 
Since we again find reasonable agreements, it can be conjectured that 
there is an equivalence between the semiclassical method and random 
matrix theory. 
\par
In the case of the GOE-GUE crossover, the corresponding 
random matrix model can analytically be treated\cite{weidenmuller} 
and the results can be compared with the semiclassical 
formulas\cite{haake}. It seems possible to apply similar 
techniques to the GOE-GSE crossover. For example, the 
diagrammatic perturbation theory is able to give the leading terms 
of the transport properties\cite{brouwer3,cremers,beri1,beri2}. 
It would be interesting to compare the semiclassical formulas with 
such analytical results and confirm the equivalence mentioned above. 

\begin{figure}[ht]
\epsfxsize=9cm
\centerline{\epsfbox{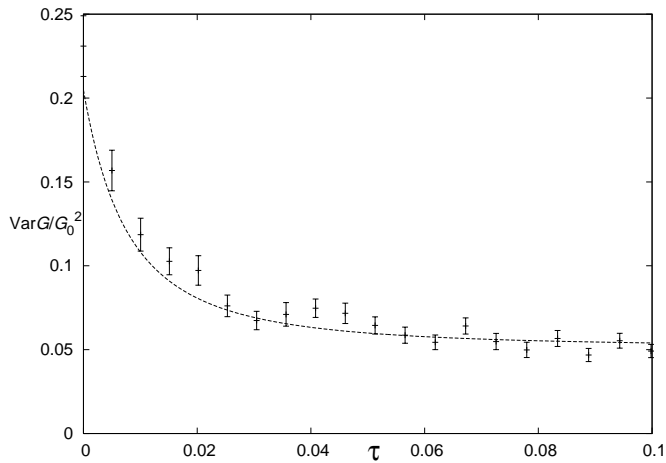}}
\caption{A comparison of the semiclassical result (curve) and 
numerically calculated random matrix results (errorbars) for 
${\rm Var}G/G_0^2$.}
\end{figure}

\begin{figure}[ht]
\epsfxsize=9cm
\centerline{\epsfbox{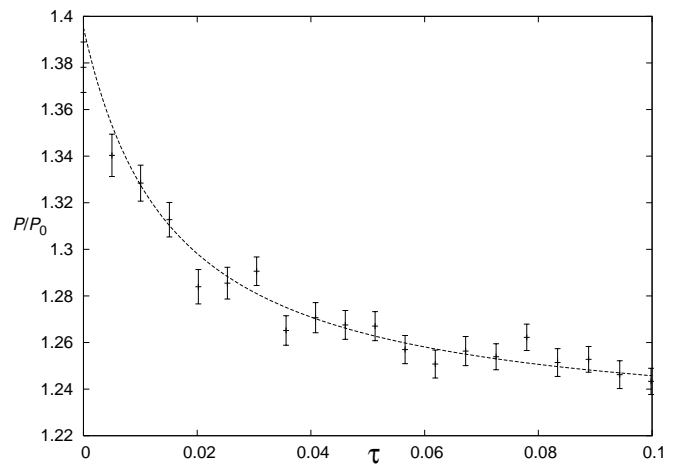}}
\caption{A comparison of the semiclassical result (curve) and 
numerically calculated random matrix results (errorbars) for 
$P/P_0$.}
\end{figure}

\section{Summary}

We studied a chaotic quantum transport of an electron with spin-orbit 
interaction in a cavity. Our approach is based on the 
semiclassical theory. The key ingredient of the theory is the 
universal statistics of the stability amplitudes. The electron 
diffusion in the position and momentum space is related to 
the escape rate, and the spin diffuses on the Bloch sphere due 
to the spin-orbit interaction, where the momentum variable plays 
the role of a stochastic magnetic field. Consequently the crossover 
parameter depends on the diffusion constant of the spin.
The spin diffusion terms appear in a non-Abelian way along the 
classical trajectories. The spins along the trajectories
interfere with each other, resulting in the change of the total 
spin. For instance, Eq.~(\ref{instance}) has both singlet and 
triplet contributions. These kinds of interference effects seem to 
play a crucial role. 
\par
In our calculation of the physical quantities such as the average 
conductance, only the first several terms of the resulting expansions 
were worked out. Such truncated results are valid only 
when the channel numbers are large. In order to obtain the full 
expansions, a more systematic calculation of the spin diffusion terms would be 
necessary. Although our expansions are truncated, they are still useful 
in the experimental point of view, because the channel 
numbers can be very large. Moreover the GOE-GSE crossover can be realized 
when the spin-orbit interaction is controlled by applying an electric field. 
Therefore we believe that an experimental test of our theory 
is in principle possible. 
\par
In addition to the transport properties analyzed in this paper, 
the shot-noise variance is also an important quantity, which 
can be treated in an RMT framework\cite{savin08}. 
It seems possible to apply the semiclassical method to the 
shot-noise variance. More ambitiously, one might be able to 
calculate arbitrary order cumulants of the conductance and shot 
noise in a semiclassical framework, as the RMT approach has 
already made a progress in that direction\cite{savin09}.
\par
Our theory is valid in the case that the dwell 
time of the electron is much larger than the Ehrenfest time. 
When the Ehrenfest time is relatively large, the resulting 
corrections should be considered\cite{brouwer4}. In addition, 
the spin diffusion mechanism might depend on the specific form of 
the spin-orbit coupling\cite{cremers,richter2}. These problems are 
also interesting in experimentally realizable situations.

\section*{Acknowledgements}
KS thanks Dr. Shiro Kawabata for a useful discussion. The authors 
are grateful to Dr. Benj\'amin B\'eri for drawing their attention to 
related studies\cite{beri1,beri2}. This work is partially supported by 
Japan Society for the Promotion of Science (KAKENHI 20540372).
\bibliographystyle{prsty}

\begin{thebibliography}{10}
\bibitem{comment}
Therefore the formulas for the average conductance and shot noise
are different from the standard expressions (See, e.g., Ref.\onlinecite{savin})
by a factor $2$, and the formula for the conductance variance 
is different by a factor $4$.
\end{thebibliography}

\begin{widetext}
\appendix
\section{Spin diffusion terms}

In this Appendix, we present the spin diffusion terms 
contributing to the average conductance, conductance variance 
and shot noise. Here $T_j$ and $t_j$ 
are the times elapsed on the loop $L_j$ and encounter $E_j$, 
respectively, and $a = \gamma_{\rm so}^2 {\cal D}$. These 
spin diffusion terms were evaluated by using Mathematica.  

\subsection{Average conductance}

The spin diffusion terms contributing to the average conductance 
(corresponding to the diagrams shown in Fig.2) are listed below.
\begin{flalign}
{\rm (i):} \ & \left\langle\!\left\langle
{\rm Tr} \left\{  
\Delta_{L_1} \Delta_{E_1} \Delta_{L_2} \Delta_{E_2} \Delta_{L_3}
\Delta_{E_1} \Delta_{L_4} \Delta_{E_2} \Delta_{L_5} 
\left( \Delta_{L_1} \Delta_{E_1} \Delta_{L_4} \Delta_{E_2} \Delta_{L_3}
\Delta_{E_1} \Delta_{L_2} \Delta_{E_2} \Delta_{L_5} \right)^{\dagger}
\right\} \right\rangle\!\right\rangle & \nonumber \\ 
 & = \left\langle\!\left\langle
{\rm Tr} \left\{ 
 \Delta_{L_2} \Delta_{E_2} \Delta_{L_3}
\Delta_{E_1} \Delta_{L_4}  
(\Delta_{L_2})^{-1} (\Delta_{E_1})^{-1} (\Delta_{L_3})^{-1} 
(\Delta_{E_2})^{-1} (\Delta_{L_4})^{-1}
\right\} \right\rangle\!\right\rangle & \nonumber \\
 & \textstyle = 
\frac{1}{2} + \frac{3}{2} e^{-a(2 t_1 + 2 t_2 + 2 T_2 + 2 T_3 )}
+ \frac{3}{2} e^{-a( 2 T_2 + 2 T_4 )} 
+ \frac{3}{2} e^{-a( 2 t_1 + 2 t_2 + 2 T_3 + 2 T_4 )}
-3 e^{-a( 2 t_1 + 2 t_2 + 2 T_2 + 2 T_3  + 2 T_4 )}. 
 & \nonumber
\end{flalign}
\begin{flalign}
{\rm (ii):} \ & \left\langle\!\left\langle
{\rm Tr} \left[  
\Delta_{L_1} \Delta_{E_1} \Delta_{L_2} \Delta_{E_2} \Delta_{L_3}
(\Delta_{E_1})^{-1} \Delta_{L_4} \Delta_{E_2} \Delta_{L_5} 
\left\{ \Delta_{L_1} \Delta_{E_1} (\Delta_{L_3})^{-1} (\Delta_{E_2})^{-1} 
(\Delta_{L_4})^{-1} \Delta_{E_1} \Delta_{L_2} \Delta_{E_2} \Delta_{L_5} \right\}^{\dagger}
\right] \right\rangle\!\right\rangle & \nonumber \\
 & = \left\langle\!\left\langle
{\rm Tr} \left\{  
 \Delta_{L_2} \Delta_{E_2} \Delta_{L_3}
(\Delta_{E_1})^{-1} \Delta_{L_4}  
(\Delta_{L_2})^{-1} (\Delta_{E_1})^{-1} \Delta_{L_4}
\Delta_{E_2} \Delta_{L_3}
\right\} \right\rangle\!\right\rangle & \nonumber \\
 & \textstyle =  
\frac{1}{2} - \frac{3}{2} e^{-a( 2 t_2 + 2 T_2 + 2 T_3 )}
- \frac{3}{2} e^{-a( 2 t_1 + 2 T_2 + 2 T_4 )} 
+ \frac{3}{2} e^{-a( 2 t_1 + 2 t_2 + 2 T_3 + 2 T_4 )}
+ 3 e^{-a(2 t_1 + 2 t_2 + 2 T_2 + 2 T_3  + 2 T_4 )}. 
 & \nonumber
\end{flalign}
\begin{flalign}
{\rm (iii):} \ & \Bigl\langle\!\Bigl\langle
{\rm Tr} \Bigl[  
\Delta_{L_1} \Delta_{E_1} \Delta_{L_2} \Delta_{E_2} \Delta_{L_3}
\Delta_{E_1} \Delta_{L_4} (\Delta_{E_2})^{-1} \Delta_{L_5}   
& \nonumber \\
 &  ~~~~~~~~~~~~~~~~~~~~~~~~~~~~~~~~~~~~~~\times
\left\{ \Delta_{L_1} \Delta_{E_1} \Delta_{L_4} (\Delta_{E_2})^{-1} 
(\Delta_{L_2})^{-1}
(\Delta_{E_1})^{-1} (\Delta_{L_3})^{-1} (\Delta_{E_2})^{-1} \Delta_{L_5} \right\}^{\dagger}
\Bigr] \Bigr\rangle\!\Bigr\rangle & \nonumber \\
 & = \left\langle\!\left\langle
{\rm Tr} \left\{  
 \Delta_{L_2} \Delta_{E_2} \Delta_{L_3}
\Delta_{E_1} \Delta_{L_4}  
\Delta_{L_3} \Delta_{E_1} \Delta_{L_2}
\Delta_{E_2} (\Delta_{L_4})^{-1}
\right\} \right\rangle\!\right\rangle & \nonumber \\
 & \textstyle =  
\frac{1}{2} - \frac{3}{2} e^{-a( 2 t_1 + 2 T_3 + 2 T_4 )}
+ \frac{3}{2} e^{-a( 2 t_1 + 2 t_2 + 2 T_2 + 2 T_3 )} 
- \frac{3}{2} e^{-a( 2 t_2 + 2 T_2 + 2 T_4 )}
+ 3 e^{-a( 2 t_1 + 2 t_2 + 2 T_2 + 2 T_3  + 2 T_4 )}. 
 & \nonumber
\end{flalign}
\begin{flalign}
{\rm (iv):} \ &  \Bigl\langle\!\Bigl\langle
{\rm Tr} \Bigl[  
\Delta_{L_1} \Delta_{E_1} \Delta_{L_2} (\Delta_{E_1})^{-1} \Delta_{L_3}
\Delta_{E_2} \Delta_{L_4} (\Delta_{E_2})^{-1} \Delta_{L_5} 
& \nonumber \\
& ~~~~~~~~~~~~~~~~~~~~~~~~~~~~~~~~~~~~~~\times
\left\{ \Delta_{L_1} \Delta_{E_1} (\Delta_{L_2})^{-1} 
(\Delta_{E_1})^{-1} \Delta_{L_3}
\Delta_{E_2} (\Delta_{L_4})^{-1} (\Delta_{E_2})^{-1} \Delta_{L_5} \right\}^{\dagger}
\Bigr] \Bigr\rangle\!\Bigr\rangle & \nonumber \\
& = \left\langle\!\left\langle
{\rm Tr} \left\{  
 \Delta_{L_2} (\Delta_{E_1})^{-1} \Delta_{L_3}
\Delta_{E_2} \Delta_{L_4}  
\Delta_{L_4} (\Delta_{E_2})^{-1} (\Delta_{L_3})^{-1} 
\Delta_{E_1} \Delta_{L_2}
\right\} \right\rangle\!\right\rangle & \nonumber \\
& \textstyle =
\frac{1}{2} - \frac{3}{2} e^{-a 2 T_2 }
+ \frac{9}{2} e^{-a(  2 T_2 + 2T_4 )}
- \frac{3}{2} e^{-a 2T_4 }. 
 & \nonumber 
\end{flalign}
\begin{flalign}
{\rm (v):} \ & \Bigl\langle\!\Bigl\langle
{\rm Tr} \Bigl[   
\Delta_{L_1} \Delta_{E_1} \Delta_{L_2} \Delta_{E_2} \Delta_{L_3}
(\Delta_{E_2})^{-1} \Delta_{L_4} (\Delta_{E_1})^{-1} \Delta_{L_5} 
& \nonumber \\
& ~~~~~~~~~~~~~~~~~~~~~~~~~~~~~~~~~~~~~\times 
\left\{ \Delta_{L_1} \Delta_{E_1} (\Delta_{L_4})^{-1} \Delta_{E_2} \Delta_{L_3}
(\Delta_{E_2})^{-1} (\Delta_{L_2})^{-1} (\Delta_{E_1})^{-1} \Delta_{L_5} \right\}^{\dagger}
\Bigr] \Bigr\rangle\!\Bigr\rangle & \nonumber \\
& = \left\langle\!\left\langle
{\rm Tr} \left\{  
 \Delta_{L_2} \Delta_{E_2} \Delta_{L_3}
(\Delta_{E_2})^{-1} \Delta_{L_4}  
\Delta_{L_2} \Delta_{E_2} (\Delta_{L_3})^{-1} 
(\Delta_{E_2})^{-1} \Delta_{L_4}
\right\} \right\rangle\!\right\rangle & \nonumber \\
& \textstyle =
\frac{1}{2} - \frac{3}{2} e^{-a 2 T_3 }
+ \frac{3}{2} e^{-a(  2 T_2 + 2T_4 )}
+ \frac{3}{2} e^{-a(  2 T_2 + 2 T_3 + 2T_4 )  }. 
 & \nonumber
\end{flalign}
\begin{flalign}
{\rm (vi):} \ & \left\langle\!\left\langle
{\rm Tr} \left[  
\Delta_{L_1} \Delta_{E_1} \Delta_{L_2} \Delta_{E_1} \Delta_{L_3}
\Delta_{E_1} \Delta_{L_4} 
\left\{ \Delta_{L_1} \Delta_{E_1} \Delta_{L_3} \Delta_{E_1} \Delta_{L_2}
\Delta_{E_1} \Delta_{L_4} \right\}^{\dagger}
\right] \right\rangle\!\right\rangle & \nonumber \\
& = \left\langle\!\left\langle
{\rm Tr} \left\{  
 \Delta_{L_2} \Delta_{E_1} \Delta_{L_3}
(\Delta_{L_2})^{-1} 
 (\Delta_{E_1})^{-1} 
(\Delta_{L_3})^{-1} \right\}  \right\rangle\!\right\rangle
 & \nonumber \\
& \textstyle =
\frac{1}{2} + \frac{3}{2} e^{-a (2 t_1 + 2 T_2)  }
+ \frac{3}{2} e^{-a(  2 t_1 + 2T_3 )}
+ \frac{3}{2} e^{-a(  2 T_2 + 2 T_3  )  }
-3 e^{-a(  2 t_1 + 2 T_2 + 2 T_3  )  }. 
 & \nonumber 
\end{flalign}
\begin{flalign}
{\rm (vii):} \ & \left\langle\!\left\langle
{\rm Tr} \left[  
\Delta_{L_1} \Delta_{E_1} \Delta_{L_2} (\Delta_{E_1})^{-1} \Delta_{L_3}
\Delta_{E_1} \Delta_{L_4} 
\left\{ \Delta_{L_1} \Delta_{E_1} (\Delta_{L_2})^{-1} (\Delta_{E_1})^{-1} 
(\Delta_{L_3})^{-1}
\Delta_{E_1} \Delta_{L_4} \right\}^{\dagger}
\right] \right\rangle\!\right\rangle & \nonumber \\
& = \left\langle\!\left\langle
{\rm Tr} \left\{  
 \Delta_{L_2} (\Delta_{E_1})^{-1} \Delta_{L_3}
\Delta_{L_3}
\Delta_{E_1}
\Delta_{L_2} \right\}  \right\rangle\!\right\rangle & \nonumber \\
& \textstyle =
\frac{1}{2} - \frac{3}{2} e^{-a  2 T_2  }
+ \frac{9}{2} e^{-a(  2 T_2 + 2 T_3 )}
- \frac{3}{2} e^{-a  2 T_3  }. 
 & \nonumber
\end{flalign}
\begin{flalign}
{\rm (viii):} \ & \left\langle\!\left\langle
{\rm Tr} \left[  
\Delta_{L_1} \Delta_{E_1} \Delta_{L_2} \Delta_{E_1} \Delta_{L_3}
(\Delta_{E_1})^{-1} \Delta_{L_4} 
\left\{ \Delta_{L_1} \Delta_{E_1} \Delta_{L_3} (\Delta_{E_1})^{-1} 
(\Delta_{L_2})^{-1}
(\Delta_{E_1})^{-1} \Delta_{L_4} \right\}^{\dagger}
\right] \right\rangle\!\right\rangle & \nonumber \\
& = \left\langle\!\left\langle
{\rm Tr} \left\{  
 \Delta_{L_2} \Delta_{E_1} \Delta_{L_3}
\Delta_{L_2}
\Delta_{E_1}
(\Delta_{L_3})^{-1} \right\} \right\rangle\!\right\rangle
 & \nonumber \\
& \textstyle =
\frac{1}{2} + \frac{3}{2} e^{-a  ( 2 t_1 + 2 T_2)   }
+ \frac{3}{2} e^{-a( 2 t_1 + 2 T_2 + 2 T_3 )}
- \frac{3}{2} e^{-a  2 T_3  }. 
 & \nonumber
\end{flalign}
\begin{flalign}
{\rm (ix):} \ & \left\langle\!\left\langle
{\rm Tr} \left[  
\Delta_{L_1} \Delta_{E_1} \Delta_{L_2} (\Delta_{E_1})^{-1} \Delta_{L_3}
(\Delta_{E_1})^{-1} \Delta_{L_4} 
\left\{ \Delta_{L_1} \Delta_{E_1} (\Delta_{L_3})^{-1} \Delta_{E_1}
\Delta_{L_2} (\Delta_{E_1})^{-1} \Delta_{L_4} \right\}^{\dagger}
\right] \right\rangle\!\right\rangle & \nonumber \\
& = \left\langle\!\left\langle
{\rm Tr} \left\{  
 \Delta_{L_2} (\Delta_{E_1})^{-1} \Delta_{L_3}
(\Delta_{L_2})^{-1}
(\Delta_{E_1})^{-1}
\Delta_{L_3} \right\} \right\rangle\!\right\rangle
 & \nonumber \\
& \textstyle =
\frac{1}{2} - \frac{3}{2} e^{-a  2 T_2   }
+ \frac{3}{2} e^{-a( 2 t_1 + 2 T_3 )}
+ \frac{3}{2} e^{-a ( 2 t_1 + 2 T_2 +  2 T_3 )}. 
 & \nonumber
\end{flalign}

\subsection{Conductance variance}

The spin diffusion terms contributing to the conductance variance 
(corresponding to the diagrams shown in Fig.3) are listed below.
\begin{flalign}
{\rm (i):} \ & \left\langle\!\left\langle
 {\rm Tr}\left[ \Delta_{L_1} \Delta_{E_1} \Delta_{L_2} (\Delta_{E_1})^{-1} \Delta_{L_3}
\left\{ \Delta_{L_1} \Delta_{E_1} (\Delta_{L_2})^{-1} (\Delta_{E_1})^{-1} \Delta_{L_3}
\right\}^{\dagger} \right] {\rm Tr}\left\{ \Delta_{L_4} (\Delta_{L_4})^{\dagger} \right\} 
\right\rangle\!\right\rangle & \nonumber \\
& = \left\langle\!\left\langle 2 \ {\rm Tr}\left\{ (\Delta_{L_2})^2 \right\} 
\right\rangle\!\right\rangle 
= 
-2 + 6 e^{-a 2T_2}. 
 & \nonumber
\end{flalign}
\begin{flalign}
{\rm (ii):} \ & \left\langle\!\left\langle 
{\rm Tr} \left\{ \Delta_{L_1} \Delta_{E_1} \Delta_{L_2} \Delta_{E_2} \Delta_{L_3}
\left( \Delta_{L_1} \Delta_{E_1} \Delta_{L_5} \Delta_{E_2} \Delta_{L_3}
\right)^{\dagger}
\right\} {\rm Tr} \left\{ 
\Delta_{L_4}\Delta_{E_1}\Delta_{L_5} \Delta_{E_2} \Delta_{L_6} 
\left( \Delta_{L_4} \Delta_{E_1} \Delta_{L_2}\Delta_{E_2}\Delta_{L_6} \right)^{\dagger}
\right\} \right\rangle\!\right\rangle & \nonumber \\
& = \left\langle\!\left\langle \left| {\rm Tr} \left\{ \Delta_{L_2} (\Delta_{L_5})^{\dagger} 
\right\} \right|^2 \right\rangle\!\right\rangle = 
1 + 3 e^{-a(2 T_2 + 2 T_5 )}.
 & \nonumber
\end{flalign}
\begin{flalign}
{\rm (iii):} \ & 
\left\langle\!\left\langle{\rm Tr} \left\{ 
\Delta_{L_1}\Delta_{E_1}\Delta_{L_2}\Delta_{E_2}\Delta_{L_3}
\Delta_{E_1}\Delta_{L_4} \left( 
\Delta_{L_1}\Delta_{E_1}\Delta_{L_4} \right)^{\dagger} \right\}
{\rm Tr} \left\{ \Delta_{L_5}\Delta_{E_2}\Delta_{L_6} 
\left(\Delta_{L_5}\Delta_{E_2}\Delta_{L_3}\Delta_{E_1}\Delta_{L_2}\Delta_{E_2}\Delta_{L_6}\right)^{\dagger} \right\}
\right\rangle\!\right\rangle & \nonumber \\
& = \left\langle\!\left\langle \left| {\rm Tr} (\Delta_{L_2} \Delta_{E_2} \Delta_{L_3}
\Delta_{E_1}) \right|^2 \right\rangle\!\right\rangle
= 
1 + 3 e^{-a( 2 t_1 + 2 t_2 + 2 T_2 + 2 T_3)}.
 & \nonumber
\end{flalign}
\begin{flalign}
{\rm (iv):} \ & 
\left\langle\!\left\langle {\rm Tr} 
\left\{ \Delta_{L_1} \Delta_{E_1} \Delta_{L_2} \Delta_{E_1} \Delta_{L_3}
\left(\Delta_{L_1}\Delta_{E_1}\Delta_{L_3}\right)^{\dagger} \right\} 
 {\rm Tr} \left\{  \Delta_{L_4} \Delta_{E_1} \Delta_{L_5}
\left(\Delta_{L_4} \Delta_{E_1} \Delta_{L_2} \Delta_{E_1}\Delta_{L_5} 
\right)^{\dagger} \right\}
\right\rangle\!\right\rangle & \nonumber \\
& = \left\langle\!\left\langle \left| 
{\rm Tr} (\Delta_{L_2} \Delta_{E_1}) \right|^2 
\right\rangle\!\right\rangle
= 
1 + 3 e^{-a (2 t_1 + 2 T_2 )}. 
 & \nonumber 
\end{flalign}
\begin{flalign}
{\rm (v):} \ & 
\left\langle\!\left\langle{\rm Tr}
\left[ \Delta_{L_1}\Delta_{E_1} \Delta_{L_2} \Delta_{E_2}\Delta_{L_3} 
\left\{ \Delta_{L_1} \Delta_{E_1} (\Delta_{L_5})^{-1} \Delta_{E_2} 
\Delta_{L_3} \right\}^{\dagger}
\right] \right. \right. & \nonumber \\
& \times \left. \left. 
{\rm Tr} \left[
\Delta_{L_4} (\Delta_{E_2})^{-1} \Delta_{L_5} (\Delta_{E_1})^{-1}
\Delta_{L_6} 
\left\{ \Delta_{L_4} (\Delta_{E_2})^{-1} 
(\Delta_{L_2})^{-1} (\Delta_{E_1})^{-1} \Delta_{L_6}
\right\}^{\dagger}
\right]
 \right\rangle\!\right\rangle & \nonumber \\
& = \left\langle\!\left\langle \left\{ {\rm Tr} 
(\Delta_{L_2} \Delta_{L_5}) \right\}^2 \right\rangle\!\right\rangle
= 1 + 3 e^{-a (2 T_2 + 2 T_5 )}.
 & \nonumber 
\end{flalign}
\begin{flalign}
{\rm (vi):} \ & 
\left\langle\!\left\langle {
\rm Tr} \left\{\Delta_{L_1}\Delta_{E_1}\Delta_{L_2}\Delta_{E_2}\Delta_{L_3}
\Delta_{E_1} \Delta_{L_4} \left(\Delta_{L_1} \Delta_{E_1} \Delta_{L_4} 
\right)^{\dagger} \right\} \right. \right. & \nonumber \\
& \times \left. \left. 
{\rm Tr} \left[ 
\Delta_{L_5} (\Delta_{E_2})^{-1} \Delta_{L_6} 
\left\{ \Delta_{L_5} (\Delta_{E_2})^{-1} (\Delta_{L_2})^{-1} (\Delta_{E_1})^{-1}
(\Delta_{L_3})^{-1} (\Delta_{E_2})^{-1} \Delta_{L_6} \right\}^{\dagger}
\right]
 \right\rangle\!\right\rangle & \nonumber \\
& = \left\langle\!\left\langle \left\{ {\rm Tr} (\Delta_{L_2} \Delta_{E_2} 
\Delta_{L_3}\Delta_{E_1})  \right\}^2 \right\rangle\!\right\rangle
= 
1 + 3 e^{-a( 2 t_1 + 2 t_2 + 2 T_2 + 2 T_3 )}.
 & \nonumber 
\end{flalign}
\begin{flalign}
{\rm (vii):} \ & 
\left\langle\!\left\langle
{\rm Tr} \left\{ 
\Delta_{L_1} \Delta_{E_1}\Delta_{L_2} \Delta_{E_1} \Delta_{L_3} 
\left(\Delta_{L_1}\Delta_{E_1} \Delta_{L_3}\right)^{\dagger}
\right\} {\rm Tr} \left[
\Delta_{L_4} (\Delta_{E_1})^{-1} \Delta_{L_5} 
\left\{ \Delta_{L_4} (\Delta_{E_1})^{-1} 
(\Delta_{L_2})^{-1} (\Delta_{E_1})^{-1} \Delta_{L_5}
\right\}^{\dagger}
\right]
\right\rangle\!\right\rangle & \nonumber \\
& = \left\langle\!\left\langle 
\left\{ {\rm Tr} (\Delta_{L_2} \Delta_{E_1}) 
  \right\}^2 \right\rangle\!\right\rangle
= 1 + 3 e^{-a( 2 t_1 + 2 T_2 )}. 
 & \nonumber
\end{flalign}
\begin{flalign}
{\rm (viii):} \ & 
\left\langle\!\left\langle {\rm Tr} \left\{ 
\Delta_{L_1}\Delta_{E_1}\Delta_{L_2} \left( 
\Delta_{L_1}\Delta_{E_1}\Delta_{L_4} \right)^{\dagger} \right\} 
{\rm Tr} \left\{ 
\Delta_{L_3}\Delta_{E_1}\Delta_{L_4} \left( 
\Delta_{L_3} \Delta_{E_1}\Delta_{L_2} \right)^{\dagger} \right\}
\right\rangle\!\right\rangle & \nonumber \\
& = \left\langle\!\left\langle 
\left| {\rm Tr} (\Delta_{L_2} \Delta_{L_4}^{\dagger})  
  \right|^2 \right\rangle\!\right\rangle
= 1 + 3 e^{-a(2 T_2 + 2 T_4)}. 
 & \nonumber
\end{flalign}

\subsection{Shot noise}

The spin diffusion terms contributing to the shot noise 
(corresponding to the diagrams shown in Fig.4) are listed below.
\begin{flalign}
{\rm (i):} \ & 
\left\langle\!\left\langle {\rm Tr} \left[  
\Delta_{L_1}\Delta_{E_1}\Delta_{L_2} \left( 
\Delta_{L_3}\Delta_{E_1} \Delta_{L_2} \right)^{\dagger}
\Delta_{L_3}\Delta_{E_1}\Delta_{L_4} \Delta_{E_2}\Delta_{L_5} (\Delta_{E_2})^{-1} 
\Delta_{L_6} \right. \right. \right. & \nonumber \\ 
& \times \left. \left. \left. 
\left\{ \Delta_{L_1} \Delta_{E_1}\Delta_{L_4} \Delta_{E_2} 
(\Delta_{L_5})^{-1} (\Delta_{E_2})^{-1}
\Delta_{L_6} \right\}^{\dagger} \right]
\right\rangle\!\right\rangle 
 & \nonumber \\
& = \left\langle\!\left\langle 
{\rm Tr} \left\{  
 (\Delta_{L_5} )^2 \right\} 
\right\rangle\!\right\rangle 
 = -1 + 3 e^{-2a T_5}.
 & \nonumber 
\end{flalign}
\begin{flalign}
{\rm (ii):} \ & 
\left\langle\!\left\langle 
{\rm Tr} \left[ \Delta_{L_1}\Delta_{E_1}\Delta_{L_2} 
\left\{ \Delta_{L_5} \Delta_{E_2} (\Delta_{L_3})^{-1} \Delta_{E_1} \Delta_{L_2}
\right\}^{\dagger}
\Delta_{L_5} \Delta_{E_2} \Delta_{L_4} (\Delta_{E_1})^{-1}\Delta_{L_3} (\Delta_{E_2})^{-1}
\Delta_{L_6} \right. \right. \right. & \nonumber \\
& \times \left. \left. \left.  
\left\{ \Delta_{L_1} \Delta_{E_1} (\Delta_{L_4})^{-1} 
(\Delta_{E_2})^{-1} \Delta_{L_6} \right\}^{\dagger}
\right]
\right\rangle\!\right\rangle & \nonumber \\
& = \left\langle\!\left\langle 
{\rm Tr}\left\{ 
 ( \Delta_{E_1}^{\dagger} \Delta_{L_3}\Delta_{L_4} )^2 \right\} 
\right\rangle\!\right\rangle 
= 
-1 + 3 e^{-2a( t_1 + T_3 + T_4 )}.
 & \nonumber 
\end{flalign}
\begin{flalign}
{\rm (iii):} \ & 
\left\langle\!\left\langle 
{\rm Tr} \left[
\Delta_{L_1} \Delta_{E_1} \Delta_{L_2}
\left(\Delta_{L_3} \Delta_{E_1} \Delta_{L_2} \right)^{\dagger}
\Delta_{L_3}\Delta_{E_1} \Delta_{L_4} (\Delta_{E_1})^{-1} \Delta_{L_5}
\left\{ \Delta_{L_1} \Delta_{E_1} (\Delta_{L_4})^{-1} (\Delta_{E_1})^{-1} \Delta_{L_5}
\right\}^{\dagger}
\right]
\right\rangle\!\right\rangle 
 & \nonumber \\
& =
\left\langle\!\left\langle {\rm Tr}\left\{  
 ( \Delta_{L_4} )^2 \right\} \right\rangle\!\right\rangle
= -1 + 3 e^{-2a T_4 }. 
 & \nonumber
\end{flalign}

\end{widetext}
\end{document}